# Imaging Intermediate Melting Phases of Dual Magnetic-Field-Stabilized Wigner Crystals

Chaofei Liu[1,#], Jianwang Zhou[1,#], Wenao Liao[1,#], Zeyu Jiang[2], Chao Zhang[1], Tingfei Guo[1], Tianyou Zhai[3], Wenhao Zhang[1,4], Ying-Shuang Fu[1,4,†], and Qi-Kun Xue[5,‡]

[1]School of Physics and Wuhan National High Magnetic Field Center, Huazhong University of Science and Technology, Wuhan 430074, China
[2]Beijing Computational Science Research Center, Beijing 100093, China
[3]State Key Laboratory of Materials Processing and Die & Mould Technology, and School of Materials Science and Engineering, Huazhong University of Science and Technology, Wuhan 430074, China
[4]Wuhan Institute of Quantum Technology, Wuhan 430206, China
[5]Department of Physics and Guangdong Basic Research Center of Excellence for Quantum Science, Southern University of Science and Technology, Shenzhen 518055, China
[#]These authors contributed equally to this work.

[†]yfu@hust.edu.cn

[‡]xueqk@sustech.edu.cn

**The competition between Coulomb repulsion and kinetic energy in correlated systems can allow electrons to crystallize into Wigner solids[1]. Despite researches across diverse two-dimensional Wigner platforms[2-10], the microscopic melting processes through possible intermediate phases remains largely unknown. Here, we present the visualization of electron-lattice melting in monolayer $VCl_3$ on graphite, where two Wigner crystals coexist with markedly different critical temperatures $T_c$ and lattice periods as stabilized by high magnetic field. One Wigner crystal possesses both record-high $T_c$ and electron density, and undergoes melting through an intermediate nematic phase upon decreasing magnetic field. In contrast, the other Wigner crystal with a lower $T_c$ yields a different intermediate phase during melting, exhibiting an anomalous electron liquid with an energy-independent modulation period. First-principles calculations corroborate the band-selective occupations of interface-transferred electrons in the formation of dual Wigner crystals. Our atomically resolved intermediate phases provide crucial insights into the microscopic melting pathways of Wigner crystals, enabling a phase diagram parameterized by both quantum and thermal fluctuations.**

The delicate interplay between Coulomb interactions and kinetic energy governs the emergence of diverse strongly correlated quantum phenomena. A paradigmatic example is the Wigner crystal, in which electrons can spontaneously form an ordered lattice under the dominant long-range Coulomb repulsion[1,11,12]. Evidences of Wigner crystallization have been observed in a broad range of two-dimensional (2D) electron systems, from traditional liquid-helium surfaces[13], GaAs/(Al,Ga)As quantum wells[2-5], to recent 2D van der Waals materials, such as graphene[6], transitional metal dichalcogenides[14,15] and their moiré superlattices[7-10]. The detections of Wigner crystals were mainly based on the insulating behavior of the crystallized carriers[3,7] or the umklapp resonances[15,16] via ensemble-averaged techniques, including transport[2], microwave impedance[9], and optical spectroscopy[2,7,8,14,15,17]. Real-space imaging of such fragile correlated crystals is challenging due to their requirements for both intricate formation conditions and non-invasive probing. Improvements in scanning tunneling microscopy (STM) technique adapted to micro-device measurements have enabled the visualization of Wigner lattices with atomic precision under electrostatic control[6,18,19]. Despite these advances, key questions regarding how the electrons crystallize remain debated, mainly over whether the electron melting is a first-order process[20,21] or a continuous transition[22-28]. For example, long-range Coulomb interaction has been proposed to forbid the first-order transition, which instead frustrates the phase separation with intermediate phases during quantum melting[22,25]. Direct imaging of possible lattice-scale intermediate phases will be crucial for a deeper understanding of the evolutionary nature of the Wigner crystal.

To address this issue, it is pivotal to establish a Wigner crystal system that is robust against the tip perturbation from STM imaging. Meanwhile, high electron density is highly desirable to regulate spin correlations therein[23,29] so as to allow the quantum-melting process of Wigner crystal via tuning magnetic field, i.e. a delicate control knob of the electron crystal[2,10,19] without alteration in the period or symmetry. Here, we employ *in situ* STM to resolve the long-sought intermediate phases at atomic precision in monolayer $VCl_3$ on graphite, where interfacial electron transfer into the heavy band of VCl$_3$ triggers robust Wigner crystal with both record-high $T_c$ and electron density. Surprisingly, two distinct Wigner crystals with different $T_c$ coexist at high magnetic fields. With decreasing magnetic field toward quantum melting, an intermediate phase with stringent nematicity is progressively developed in the higher-$T_c$ Wigner crystal. Concomitantly, the lower-$T_c$ Wigner crystal also melts into another intermediate phase of anomalous liquid. Those intermediate phases are further complemented with thermal-melting experiments, cooperatively mapping out the critical stages in the phase diagram of Wigner liquid–solid transition.

## Atomic and electronic structures of monolayer VCl$_3$

The STM experiments were performed mainly at 1.6 K (unless specified), and the VCl$_3$ thin films were grown on graphite *in situ* by molecular beam epitaxy (see Methods). VCl$_3$ is a van der Waals material with rhombohedral structure (space group: $R\bar{3}$)[30] (Fig. 1a). Within each Cl-V-Cl trilayer, the honeycomb V lattice is bonded to the bilateral two Cl sublayers (lower panel of Fig. 1a), with the V atoms octahedrally coordinated with Cl atoms. Figure 1b presents a large-scale topography (150×150 nm$^2$) of monolayer (5.1-Å thick) VCl$_3$ typical of the high crystalline quality. The magnified STM image displays the triangularly arranged clustered protrusions (Fig. 1c) owing to the trimerization of Cl atoms. Fast Fourier transformation (FFT) of the surface Cl lattice (Fig. 1d) shows sharp Bragg points of the atomic V lattice, together with the diffraction points expected for the triangular trimer lattice. Analysis of the reciprocal Bragg points yields the lattice constants of $a=b=6.0$ Å, in agreement with previous first-principles calculations[31,32].

The tunnelling d$I$/d$V$ spectra, proportional to the local density of states, reveal a global insulating gap (Fig. 1e). Such a gap arises from the strong on-site Coulomb repulsion, which splits the V-$t_{2g}$ manifolds into upper and lower Hubbard bands, dominated by the nearly flat $d_{z^2}$ and hybridized ($d_{xy}$, $d_{x^2-y^2}$+Cl-$p$) orbitals, respectively (Fig. 1f)[33]. The gap magnitude, extracted from a logarithmic plot of the d$I$/d$V$ spectrum, is 1.1 eV (inset of Fig. 1e). This value is comparable with our density functional theory plus Hubbard $U$ (DFT+$U$) calculations (1.3 eV in Fig. 1f; see Methods for details) and previous theoretical studies[33]. Beyond the gap, the remarkably flat conduction band yields an effective electron mass of $m_e^*=4.48m_0$ (e.g. along Γ-M direction; $m_0$, mass of free electron). When interfaced with graphene, VCl$_3$ would in theory possess an electron density of $n_e=2.59\times10^{13}$ cm$^{-2}$ from charge transfer. Further using an effective dielectric constant $\varepsilon_r=1.85$, which is half that of the combination of vacuum and graphite, we estimate the dimensionless interaction parameter[34] $r_s = m_e^* e^2/(4\pi\varepsilon\hbar^2\sqrt{\pi n_e})\approx50.8$ (see Methods section 'DFT estimation of $r_s$' for more details). Here, $r_s$ quantifies the ratio between Coulomb interaction and kinetic energy. Note that quantum Monte Carlo simulations[23,34,35] have shown that the 2D Wigner crystal is stabilized beyond the critical interaction parameter $r_s\approx31$. Our calculated large $r_s$ thus specifies VCl$_3$ as a promising system for STM exploration of the emergent Wigner physics.

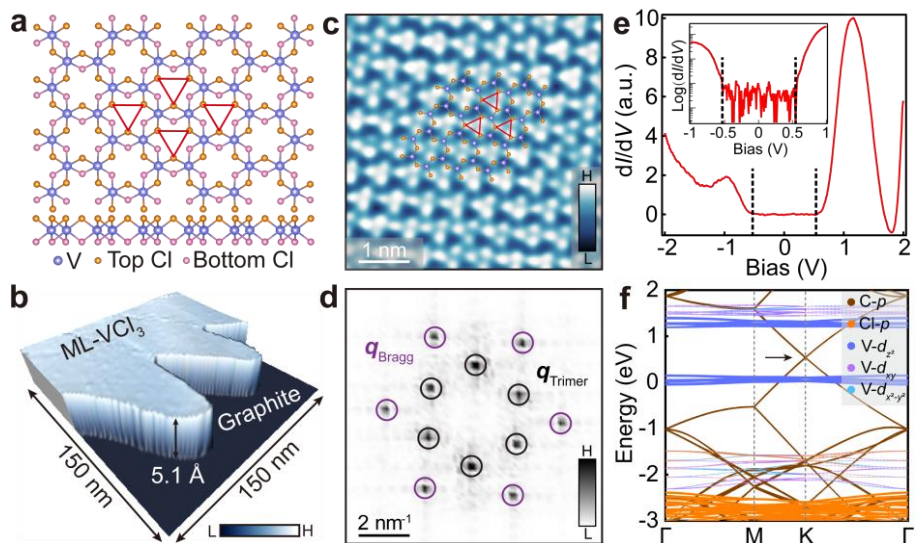

**Fig. 1| Atomic characterizations of monolayer VCl$_3$. a,** Crystal structure of VCl$_3$. **b,** Large-scale STM image of ML (monolayer) VCl$_3$ film grown on graphite ($V_s$=1.8 V, $I_t$=5 pA). **c,d,** Magnified STM image and FFT pattern ($V_s$=0.5 V, $I_t$=60 pA). **e,** Tunneling d$I$/d$V$ spectrum taken in a large-bias range. Inset: logarithm of d$I$/d$V$, to exemplify the extraction of gap magnitude. **f,** Calculated electronic structures of monolayer VCl$_3$ on graphene. The arrow marks the position of Dirac point.

## Atomic imaging of two coexisting Wigner crystals

Having characterized the monolayer VCl$_3$ film, we proceed to investigate the electronic phases therein shaped by enhanced correlations under a magnetic field of 12 T. In the STM images, a superlattice feature is clearly visible (Fig. 2a). Correspondingly, the FFT reveals diffraction spots at a wavevector $q_{SL}$ (Fig. 2b). After Fourier filtering of the FFT pattern by selectively reserving the $q_{SL}$ component, the STM image via inverse FFT presents more explicitly the real-space morphology of the superlattice (Fig. 2e). Quantitatively, FFT linecuts taken along the $q_{SL}$ direction at representative biases of ±1.5 V exhibit pronounced peaks for the superlattice near 0.31 nm$^{-1}$ (Fig. 2g), yielding a lattice period of 3.25 nm. Complete energy-resolved FFT linecuts along the same direction are analyzed to extract the energy dispersion of $q_{SL}$ (Fig. S1a). Remarkably, the obtained wavevector $q_{SL}$ shows no discernible energy dependence over a wide range up to ±[0.3, 2] eV (Fig. 2i), consistent with a static charge order[36].

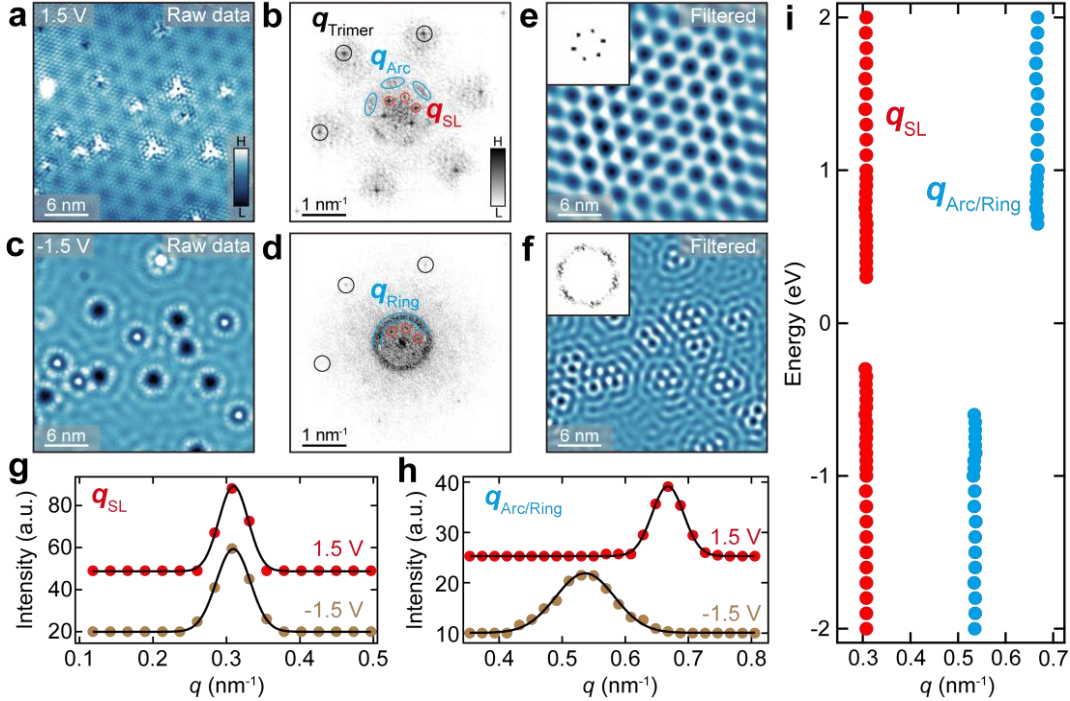

**Fig. 2| Coexistence of the charge orders with different $T_c$ under strong magnetic field. a,b,** Typical STM image of VCl$_3$ acquired at 1.5 V ($I_t$=30 pA) under a magnetic field of 12 T and the FFT pattern, showing wavevector $q_{SL}$ for the superlattice and $q_{Arc}$ for the standing wave-like feature. **c,d,** STM image of the same area as in **a**, but recorded at −1.5 V, and the FFT pattern. **e,f,** STM image via inverse FFT after Fourier filtering out the reciprocal features in **b** by reserving only the diffraction peaks of $q_{SL}$ and $q_{Arc}$, respectively. **g,h,** Plots of the FFT linecuts along $q_{SL}$ and $q_{Arc}$ directions at ±1.5 V. The solid curves represent Gaussian fittings used to determine the precise magnitude of $q_{SL}$ and $q_{Arc}$. **i,** Energy dependence of $q_{SL}$ and $q_{Arc}$ extracted from the full-bias ([−2, 2] V) plots of the FFT linecuts for d$I$/d$V$ mappings (Fig. S1).

Meanwhile, near atomic defects embedded within the film, faint standing wave-like patterns are observed (Fig. 2a), which turns more evident at negative bias (Fig. 2c). This behavior is further corroborated by the spatially resolved tunneling spectra and associated conductance near −1.0 V taken near the defect (Fig. S2), which directly reveal a periodic modulation of the spectral intensity in space. Corresponding FFT of the standing wave-like STM images consistently displays arc-like features at the wavevector $q_{Arc}$, located near the superlattice diffraction spots (Fig. 2b). With decreasing bias, the arc grows more diffusive, and evolves into an almost continuous ring at −1.5 V (Fig. 2d). The real-space morphology of pure standing wave-like pattern is better visualized in the Fourier-filtered STM image obtained by retaining only the $q_{Arc}$-related diffractions (Fig. 2f). It appears as a pre-crystallized lattice, where the defects serve as the seed of crystallization[37]. FFT linecuts along the $q_{Arc}$ direction at ±1.5 V exhibit peaks for the reciprocal arc near 0.67 and 0.54 nm$^{-1}$, respectively (Fig. 2h), corresponding to real-space periods of 1.50 and 1.85 nm. Such disparity likely arises because the tip-induced band bending that is capable of renormalizing the electronic states depends sensitively on the polarity of applied bias[38]. Due to the same reason, the arc turns more diffusive at negative bias (Fig. 2d). Its dispersion, as determined by searching the reciprocal peak in the bias-dependent FFT linecuts (Fig.

S1b), also features no energy dependence as typical of another charge order (Fig. 2i). This leaves alternative scenarios represented by quasiparticle interference (QPI) unlikely [see Methods section 'Possible explanations of the charge-density modulations'].

The observed crystallized superlattice and pre-crystallized charge orders are both hosted inside monolayer VCl$_3$, which exhibits a large $r_s$ well exceeding the threshold of $r_s \approx 31$ for Wigner crystallization. We therefore attribute these two static charge orders to dual Wigner crystals, dubbed as I and II, respectively, distinguished by strikingly different critical formation temperatures $T_c$. Further evidences supporting the identification of Wigner crystals include the orientational flexibility of the superlattice independent of the structural lattice (Fig. S3), and the collective sliding of the superlattice driven by magnetic field (Fig. S4). Both observations are inconsistent with moiré lattice or Peierls charge-density wave, and instead highlight the character of electron solids decoupled from the crystal structure.

**Intermediate phases during the melting of Wigner crystals**

Next, we study the intermediate phases emerging from the quantum-melting process of the Wigner crystals, achieved by decreasing the magnetic field[2,19], which restores the electron kinetic energy[39]. By thermal or quantum fluctuations, the Wigner crystal will be melting into a liquid, with the possibility of generating exotic intermediate phases[25,27,40] and quantum magnetism[41,42]. Microscopically, the quantum melting of Wigner crystal here is realized through the progressive development of intermediate nematic order. As shown in Fig. 3a, the Wigner crystal I (i.e., the superlattice) initially exhibits well-ordered equilateral-triangular lattice at high magnetic fields. When the magnetic field is reduced, there emerges a nematic phase with elongated-triangular lattice (bright area) coexisting with the region of equilateral-triangular lattice (dark area) (Fig. 3b-e). The onset of such electronic nematicity is accompanied by the appearance and gradual migration of domain boundary, together with the increasing proportion of nematic regions. Finally, at 0 T, the nematic phase almost fully prevails the entire surface, leaving atomically thin domain walls flanking the orientation-contrasting domains (Fig. 3f). Corresponding to the real-space images, the FFT intensity of the Wigner crystal I gradually reduces with decreasing the magnetic field (Fig. 3g-l), indicative of electron melting. Quantitatively, owing to the emergence of nematic regions, the proportion of regularly ordered Wigner crystal I decreases monotonically from 100% at ≥8 T to 13% at 0 T (Fig. 3s).

To further evaluate the intermediate phase, we zoom into a nematic domain and compare its electron lattice in the two limiting cases of high (8 T) and low (0 T) magnetic fields (Fig. 3m,o vs. n,p). As the field is removed, the nematic distortion of the Wigner lattice I becomes more explicit (Fig. 3n). Its corresponding FFT pattern exhibits a pronounced deformation, reducing the symmetry of the diffraction spots from $C_6$ at high field to $C_2$ at low field (Fig. 3o vs. p). Upon the nematic regions are fully developed at 0 T, multiple orientation-contrasting nematic Wigner crystallites from (Fig. 3f, S6), and are separated by domain boundaries. Separate inspection of these adjacent domains and their corresponding FFT images presents both pronounced shape distortion and intensity anisotropy of the diffraction patterns along different directions (Fig. S8). Despite the nematic phase, the atomic lattice of VCl$_3$ keeps unchanged (Fig. S9), demonstrating a purely electronic origin of the nematicity (Fig. S7, 10).

Decreasing the magnetic field not only triggers the nematic intermediate phase of Wigner crystal I, but also induces the quantum melting of Wigner crystal II. With decreasing magnetic field, the pre-crystallized Wigner crystal II near defects turns into an increasingly wavy-type electron liquid (Fig. 3a-f). Correspondingly, the reciprocal pattern evolves from an arc-like to a ring-like feature (Fig. 3g-l), meaning entry into the melted liquid. The liquefied behavior is better revealed by comparing the Fourier-filtered images of Wigner crystal II in the two limiting cases of contrasting fields (Fig. 3q,r). Typically, the melted Wigner crystal II appears in a form of disordered electron arrangement. This resembles the quantum densification proposed as superposition of multiple Wigner configurations during charge melting[28]. Considering its energy independence, such melted Wigner crystal is endowed an identity of anomalous electron liquid[42,43]. The field-tuned quantum melting behavior is reproducibly observed in different samples for both Wigner crystals I and II (Fig. S5).

To unveil the quantum melting mechanism, we perform quantitative analysis to the diffraction-spot intensities as a function of magnetic field. Both Wigner crystals I and II reveal suppressed diffraction intensities with decreasing magnetic field (Fig. 3s). In theory, an external field $B$ introduces Zeeman energy to the spins of charge-ordered electrons. Based on calculations using the partition function of a spin-1/2 system in the Zeeman field, the resulting magnetization is expected to follow the Brillouin function $M = \frac{1}{2}g\mu_B n \tanh\frac{g\mu_B(B-B_0)}{2k_B T}$ ($g$, Landé factor; $\mu_B$, Bohr magneton; $n$, carrier density; $k_B$, Boltzmann constant)[44]. The function well fits our experimental data of $B$-dependent diffraction intensity for both Wigner crystals after a field offset of $B_0$=2.6 T (Fig. 3s). This implies a spin-related origin of the nematicity, which is likely linked to antiferromagnetic spin stripes predicted in Wigner crystals with intermediate charge densities[45] (see Methods section 'Nematicity in the Wigner crystal' for further discussions). Within this framework, the field-driven disappearance of nematicity here arises from the polarization of in-plane spin-stripped Wigner crystallization under a sufficiently strong out-of-plane magnetic field. Furthermore, owing to the enhanced spin susceptibility of a Wigner solid compared with the liquid phase[23,34], the magnetic field yields a larger Zeeman-energy reduction for the Wigner crystal, thereby stabilizing the electron solid more effectively[43]. This explains the stronger magnetic-field stabilization of Wigner crystal I than II, manifested as a more pronounced enhancement of the diffraction intensity with increasing field (Fig. 3s).

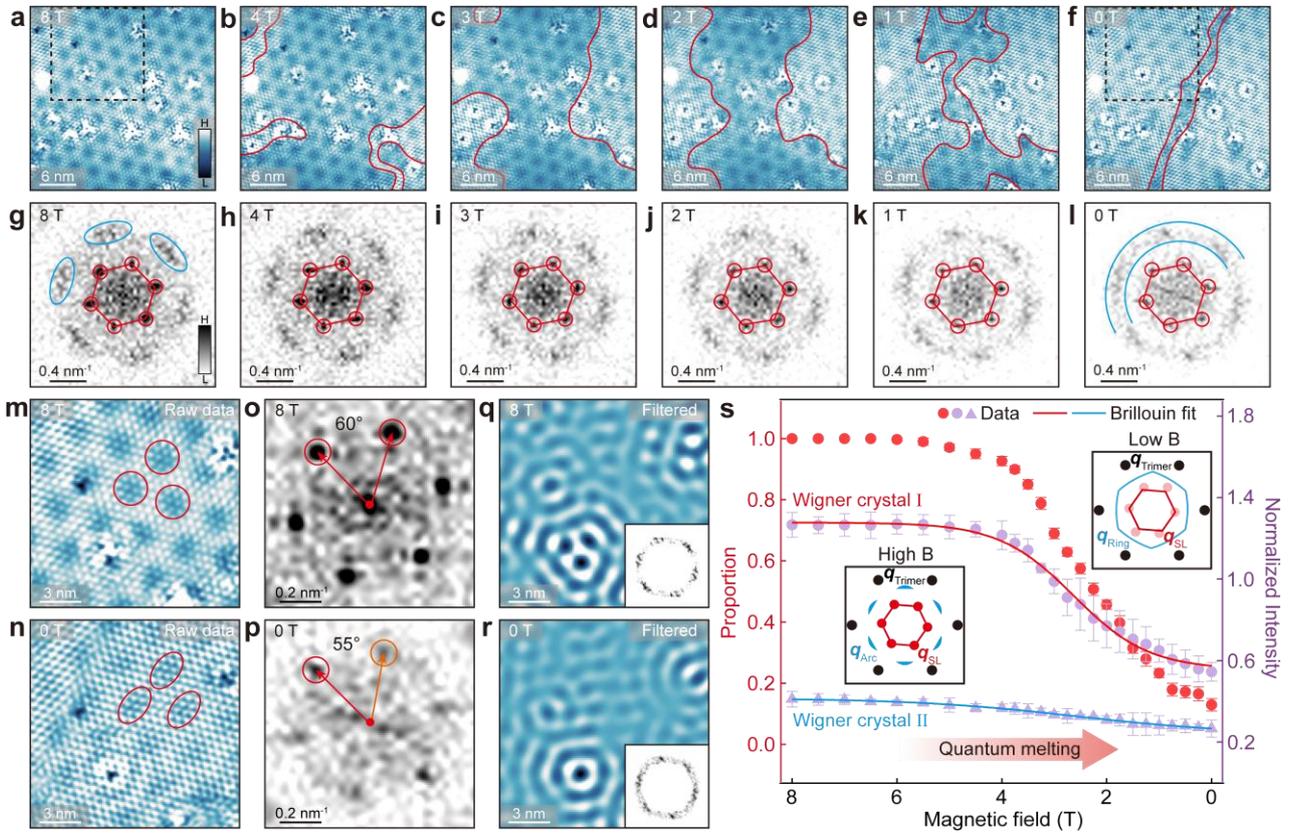

**Fig. 3| Magnetic field-tuned melting of the Wigner crystals. a-f,** STM images under selected magnetic fields, revealing the emergence of nematic-distorted Wigner lattice that appears as brighter regions, as outlined by the domain wall (red line) ($V_s$=1.5 V, $I_t$=30 pA). More detailed field evolution can be found in Fig. S11. **g-l,** FFT of the STM images in **a-f**, respectively. The regular hexagons are shown to highlight whether the distortion exists in the diffraction spots of Wigner crystal I. Note that to eliminate the field-induced structure-related factors, the FFT intensity has been normalized by that of the trimer diffraction point. **m-p,** Magnified views of the regular and distorted triangular Wigner lattices [positions boxed in **a,f**] at 8 and 0 T, respectively **m,n**, together with their FFT patterns **o,p** ($V_s$=1.5 V, $I_t$=30 pA). **q,r,** Magnified views of Fourier-filtered STM images [positions boxed in **a,f**] for pure Wigner crystal II at 8 and 0 T. **s,** Magnetic-field dependence of the proportion of regular-shaped Wigner lattice and the averaged, normalized diffraction intensity of dual Wigner crystals. The data of Wigner intensity are fitted with a Brillouin function (solid curves). The insets display the schematic reciprocal shapes of Wigner crystals I and II at representative higher and lower fields.

Thermal melting of the Wigner crystals serves as a complementarity to the microscopic picture of above quantum melting results. As increasing the temperature from 10 K to 30 K, the nematicity of Wigner crystal I is weakened with indiscernible domain walls (Fig. S12a vs. c) and reduced intensity difference in its diffraction spots (Fig. S12d vs. f). Further heating above 79 K makes the nematic order vanish completely (Fig. 4a-c), showing the diffraction spots with identical intensities and a regular $C_6$ symmetry (Fig. 4d-f). Notably, the diffraction spots of Wigner crystal I become increasingly blurred with increasing temperature, but surprisingly survive up to 190 K, i.e. the maximum temperature of our STM measurement. Similarly, the standing wave-like feature of Wigner crystal II also weakens with increasing temperature (Fig. 4a-c). Its corresponding ring-like feature in FFT images becomes blurred into a belt-like feature with pronounced intensity suppression (Fig. 4d-f). Quantitative analysis of such intensity suppression for both Wigner crystals reveals an exponential decay (Fig. 4g). The enlarging reciprocal area from ring-like to belt-like feature signifies the trend evolving from the anomalous liquid towards 'normal' Fermi liquid under increased thermal excitations.

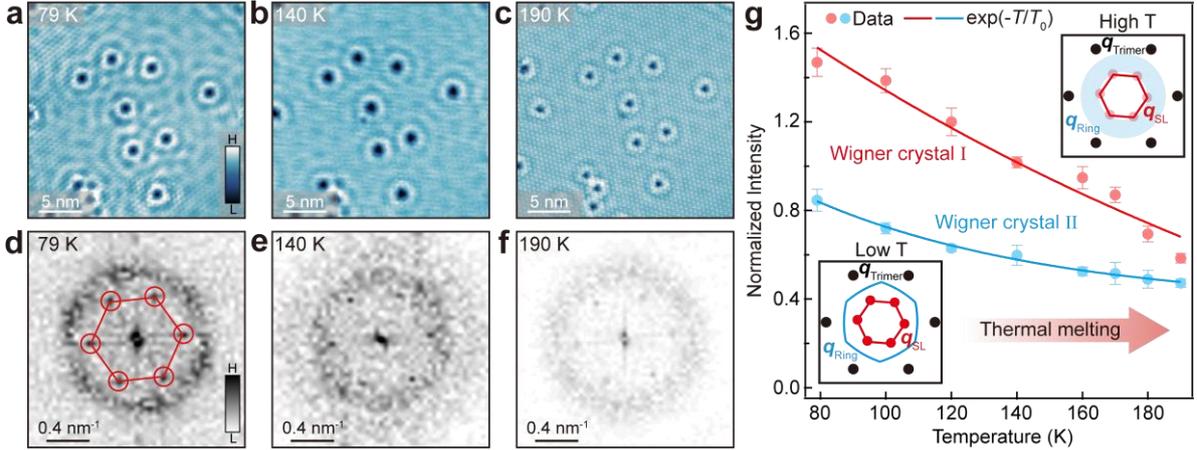

**Fig. 4| Thermal melting of the Wigner crystals at 0 T. a-f,** STM images and corresponding FFT patterns at representative temperatures ($V_s$=1 V, $I_t$=30 pA). **g,** Summarized temperature dependence of the normalized intensity for both Wigner crystals. The solid lines are fits based on an exponential decay. For complete set of STM images at viable temperatures, see Fig. S13.

## Scenario for the hierarchy of Wigner crystallizations

To explain the two distinct Wigner crystals, we propose a phenomenological model based on the coexistence of localized and itinerant electrons. The electrons forming these Wigner crystals are conjectured to originate from the transferred charges[46,47] across the interface of VCl$_3$/graphite heterojunction due to their work-function $W$ difference ($W_{VCl_3}$≈5.43 eV[48], $W_{Graphite}$≈4.4 eV[49]). Such charge transfer induces interfacial band bending, which lowers the conduction band of VCl$_3$ beneath the Fermi level, creating a reservoir of 2D electrons (Fig. 5a). The band bending recovers at the top atomic layer of VCl$_3$, making its insulating gap remain straddling the Fermi level[50]. According to our DFT+$U$ calculations using a 4×4 heterostructure supercell as an example, similar to the size of typical Wigner crystal I, ~1.3 electrons are transferred per supercell from graphene to VCl$_3$. Upon full relaxation of the doped supercell lattice, the reconstructed electronic structures feature two subbands: a nearly flat one below the Fermi level, and a more dispersive one crossing the Fermi level, derived respectively from hybridized V-3$d$/Cl-3$p$ polaronic states and delocalized V-$d_{z^2}$ orbitals (Fig. 5b; see Methods). The lower-lying nearly flat subband, which promotes strong electron localization, is fully occupied by one of the 1.3 transferred electrons. These electrons are situated in the periodically repeated supercells, with the density susceptible to extrinsic factors represented by the puddle-type inhomogeneity. They can be arranged into a Wigner crystal through a polaron-enhanced mechanism, giving rise to the observed Wigner crystal I with high $T_c$ with a variable period (see Methods section 'Distinct itinerancy for the transferred electrons' for further explanations).

The residual 0.3 electrons per supercell partially occupy the higher-energy subband with larger bandwidth, thus exhibiting a more itinerant character. They are expected to relax into a pre-crystallized electron lattice (Wigner crystal II) with a lower $T_c$ as probed experimentally. Furthermore, the Coulomb repulsion drives the transferred electrons in distinct subbands to occupy different atomic sites in real space. This naturally leads to their distinct crystallographic orientations upon Wigner crystallization, a prediction that is also supported by our inverse-FFT analysis of the two Wigner crystals (Fig. 2e vs. f). We emphasize that the phenomenological model only explains the coexisting Wigner crystals qualitatively. At a quantitative level, their exceedingly >100-fold-large $T_c$ difference and why the melted Wigner crystal II shows a nearly fixed period (Fig. S3) remain unknown. A thorough understanding would require detailed calculations by incorporating long-range Coulomb interaction, or even the inter-layer couplings beyond the short-range mean-field theories.

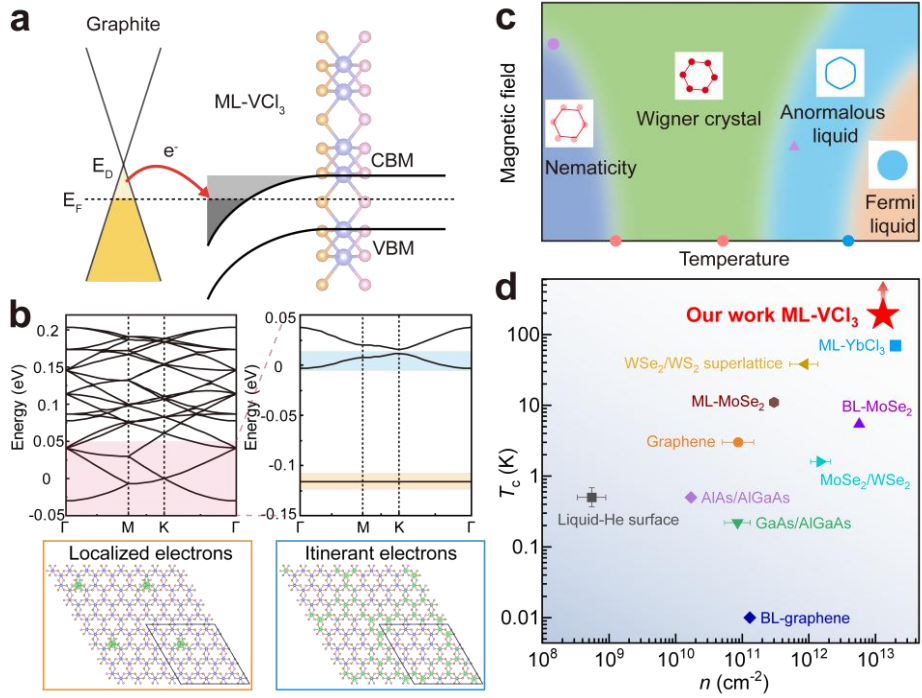

**Fig. 5| Physics picture of the Wigner crystallizations. a,** Energy band scheme of $VCl_3$/graphite heterojunction, showing the charge transfer from graphite substrate to $VCl_3$. The valance band maximum (VBM) and conduction band minimum (VBM) of $VCl_3$ accordingly bend downwards at the interface. **b,** Electronic structures of a 4×4 supercell doped with 1.3 electrons without (left panel) and with (right panel) full relaxation of the supercell lattice. Ferromagnetic ordering and DFT+$U$ approach with $U$ = 3.0 eV are applied here. The electron solid is enabled by the polarons in 4×4 supercell (left bottom), while the melted electron liquid originates from the remaining transferred electrons with stronger itinerancy (right bottom). **c,** Phase diagram presented as functions of magnetic field and temperature. The dots and triangle, color-matched to the data points in Fig. 3s and 4g, denote the estimated parameter regimes accessed in our experiments. **d,** Comparative survey of various Wigner systems with different $T_c$ and carrier density $n$ presented in logarithmic scale[2,3,5-10,13-15,18,19,28,51-54]. BL, bilayer. The arrow represents the lower $T_c$ bound of our Wigner crystal I.

**Discussions**

The lower-$T_c$ Wigner crystal II represents the fate of higher-$T_c$ Wigner crystal I at sufficiently high temperatures, which allow us to propose the procedures of Wigner melting by combining them two as follows (Fig. 5c). Initially, with increasing quantum fluctuations (i.e., decreasing magnetic field), the Wigner lattice is 'broken' into spin-correlated nematic crystallites. With then increasing thermal fluctuations, the Wigner crystallites restore the $C_6$ symmetry. The thermal crystallization behavior is promoted by destroying the in-plane magnetization with substantially enhanced spin entropy, reminiscent of the Pomeranchuk effect[55]. Such 'anomalous' crystallization has been predicted in the phase transition of Wigner crystal[24,25], and revealed in optical experiments as a rightward slant in the Wigner crystal–liquid phase boundary in $MoSe_2$ monolayer[43]. Further enhanced thermal fluctuations partly melt the crystallites into anomalous electron liquid with a nearly fixed wavelength. Upon successively increasing temperature, the fraction of the anomalous liquid gradually grows. Finally, above the melting temperature, the Fermi liquid fully develops.

The Wigner crystals here, especially the higher-$T_c$ Wigner crystal I, differ markedly from those reported earlier. To our knowledge, the Wigner crystal I, robustly surviving up to >190 K (Fig. 4f), ranks the highest reported $T_c$ of Wigner systems (Fig. 5d). Furthermore, the electron density $n_e$ of $1.1\times10^{13}$ cm$^{-2}$ inferred from $n_e=\frac{2}{\sqrt{3}a^2}$ (taking period $a$=3.25 nm) for VCl$_3$ is also surprisingly higher than that in other Wigner systems (Fig. 5d), e.g., GaAs/(Al,Ga)As quantum wells (~$10^{10}$–$10^{11}$ cm$^{-2}$), graphene (~$10^{11}$–$10^{12}$ cm$^{-2}$), and moiré superlattice (~$10^{12}$ cm$^{-2}$). Such remarkable stability can be traced to the extremely flat band occupied by the electrons constituting Wigner crystal I. The flat dispersion yields a large effective mass and, consequently, a large interaction parameter $r_s \geq 50.8$, which drives monolayer VCl$_3$ deep into the strong-correlated regime suitable for robust Wigner solid.

**Conclusion**

In summary, we have achieved atomic-scale imaging of the intermediate phases upon melting of Wigner crystals in a heterojunction of monolayer VCl$_3$/graphite. Two Wigner crystals, exhibiting distinct thermodynamic and quantum stabilities with different $T_c$, are found unexpectedly coexistent in a single material system, instead of merging into a single-period Wigner lattice. Their coexistence can be explained by a phenomenological model that incorporates both the polaron-trapped and itinerant electrons. Most importantly, the intermediate phases upon melting of higher- and lower-$T_c$ Wigner crystals are manifested as a nematic electron lattice and an anomalous electron liquid with energy-invariant wavevector, respectively. The Wigner lattice melting here is achieved via decreasing magnetic field. This strategy serves as a delicate control knob for fine-tuning the competition of Coulomb interaction and kinetic energy within the Wigner crystal, while preserving its period or symmetry. Previously, the nature of Wigner phase melting has long been hotly debated[20,21,25-27], which either predicts a first-order transition, or a sequence of intermediate solid–liquid mixed microemulsion phase[22,24,25,56,57], typically in the format of stripe- or bubble-type clusterization of electrons[25]. Our findings preclude the Wigner crystal melting as a first-order process, and suggest at microscopic level new intermediate phases different from the afore predictions. This imposes key constraints on the many-body theories describing quantum phase transitions, and are expected to stimulate future studies for establishing a universal scenario of Wigner melting.

## Methods

### STM experiments

The experiments are performed in the low-temperature STM (Unisoku 1300, 1500) equipped with a preparation chamber. The monolayer $VCl_3$ film is grown on graphite. The graphite substrate is cleaved *ex situ*, and further degassed in in vacuum at ~700 K for 2 h before growth. High-purity ultra-dry $VCl_3$ powder (99%, Alfa Aesar) is evaporated at 420 K from a home-made K-cell evaporator, and the substrate temperature is kept around 410 K during the sample growth. The base pressure is better than $8\times10^{-9}$ Torr. The STM measurement is conducted mainly at 1.6 K unless specified exclusively. An electrochemically etched W wire is used as the STM tip, which has been cleaned on the Ag(111) surface before conducting the STM experiments. The topographic images are obtained in a constant-current mode, with typical setpoints $V_b=\pm1.5$–$\pm2.0$ V, $I_t$= 30 pA. The tunneling spectra are acquired using the lock-in technique with a bias modulation of 40 mV at 983 Hz, and typical $V_b$=2V, $I_t$= 200 pA.

### DFT calculations

DFT calculations were performed using the VASP code[58] with projector augmented-wave (PAW) potentials. Throughout this work, the electronic structures were computed within the DFT+$U$ framework[59], adopting $U$=3.0 eV as in previous studies[60]. Results obtained with $U$=4.0 eV and with the hybrid Heyd-Scuseria-Ernzerhof (HSE) functional[61] were also examined to ensure that the choice of functional does not introduce qualitative changes to the proposed physics. A 4×4 supercell was used to simulate the charge transfer between $VCl_3$ and graphene and the formation of the induced polaronic state. A 12×12 $k$-point mesh was applied for unit-cell calculations (3×3 for the supercell). The plane-wave kinetic-energy cutoff was set to 400 eV, and a vacuum spacing of 20 Å was used in all calculations. Ferromagnetic order was adopted in all calculations.

### DFT estimation of $r_s$

The dimensionless parameter[62] $r_s = m_e^* e^2 / (4\pi\varepsilon\hbar^2 \sqrt{\pi n_e})$ is a key quantity for assessing the stability of the Wigner crystal. It depends on three material-specific parameters: the electron effective mass $m_e^*$, the areal electron density $n_e$, and the environmental dielectric constant $\varepsilon$. In our system, the $VCl_3$ monolayer resides in a half-screened environment: the lower half-space is screened by the graphite substrate, while the upper half-space is nearly exposed to vacuum. Therefore, it is appropriate to adopt an effective dielectric constant $\varepsilon=(1+\varepsilon_r)\varepsilon_0/2$, where $\varepsilon_r$ is the static dielectric constant of graphite. Our DFT calculation gives $\varepsilon_r$=2.7, in good agreement with previous reports on the dielectric constants[63]. Consequently, we obtain $\varepsilon$=1.85$\varepsilon_0$.

In principle, $m_e^*$ and $n_e$ should be extracted directly from the electronic structure of the $VCl_3$ monolayer and the $VCl_3$/graphite heterointerface. This, however, is nontrivial, as the electronic structure of $VCl_3$ is highly sensitive to the orbital ordering, magnetic configuration, and even the choice of substrate[33,60,64-66]. On the theoretical side, such sensitivity gives rise to a large number of (meta)stable states, making the determination of the global ground state of $VCl_3$ highly challenging. It has been shown that the lowest-energy configuration strongly depends on computational details, including the choice of exchange-correlation functional, the magnitude of the Hubbard $U$, and whether standard DFT or hybrid functionals are employed, among others. However, in our case, we are guided by experimental observations, and fortunately, we only need to estimate the order of magnitude of $r_s$ rather than determine its exact value. This allows us to bypass the complications associated with those computational details.

We must understand that all the multiple configurations of $VCl_3$ originate from two fundamental degrees of freedom, i.e., the orbital ordering and the spin ordering. Regarding the orbital degree of freedom, the strong octahedral crystal field of local $VCl_6$ cluster splits 5-fold V-3$d$ orbitals into $e_g$ and $t_{2g}$ manifolds, with the former pushes deep into the conduction band and the latter remaining near the Fermi level. Under the influence of in-plane rotational symmetry, $t_{2g}$ states further splits into $a_{1g}$ ($d_{z^2}$) and 2-fold $e'_g$ ($d_{xy}, d_{x^2-y^2}$) states. Note that in our notation, the $z$-axis is defined along the global out-of-plane direction, but not the 'natural' axis of the tilted local $VCl_6$ octahedron. The direction of $a_{1g} - e'_g$ splitting, however, is highly sensitive to the local environment and computational parameters, resulting in two (meta)stable electronic structures depending on which level lies lower in energy.

However, it has been shown that[60,65] the emergence of an $a_{1g}$-lowest-lying configuration necessarily lifts the twofold degeneracy of the $e'_g$ states, since after ionization of Cl anions, $V^{3+}$ retains only two electrons to occupy the 3$d$-orbitals. Consequently, the splitting of $e'_g$ states is expected to be accompanied by a Jahn–Teller type distortion that removes the $C_3$ rotational symmetry of the VCl$_3$ monolayer. Yet, such symmetry breaking has never been observed in our experiments, nor in previous experimental reports employing graphite substrates[66]. In fact, the complexity of determining the ground state of VCl$_3$ mainly arises from the large number of possible lattice distortions that can lift the degeneracy of the $e'_g$ states, which are energetically close to one another. By experimentally excluding the possibility of the $a_{1g}$-lowest-lying configuration, we identify the $e'_g$-lowest-lying configuration as the only viable ground state. Preserving the $C_3$ rotational symmetry, this configuration possesses no additional lattice degrees of freedom.

Next, we consider the role of the spin degree of freedom, which is likewise nontrivial, since DFT often struggles to accurately predict the relative energies of different magnetic lattices. However, our focus is not on the specific spin configuration of VCl$_3$ itself, but rather on how different spin orderings influence the kinetic energy of the doped electrons, which directly governs the formation of the Wigner crystal. In the language of a tight-binding model, the kinetic energy is determined by the effective hopping amplitude between neighboring lattice sites. The hopping is maximized when all spins are aligned corresponding to the ferromagnetic (FM) state, and minimized when adjacent sites have opposite spins, i.e., the AB-type antiferromagnetic (AB-AFM) configuration typical for a honeycomb lattice such as VCl$_3$. Therefore, it suffices to consider these two magnetic orderings, FM and AB-AFM, to capture the upper and lower bounds of the electron effective mass.

Finally, the only remaining degree of freedom is the density functional. In the literature, the electronic structure of VCl$_3$ is typically calculated using the DFT+$U$ approach with $U$ values ranging from 3.0 to 3.5 eV. To verify the robustness of the proposed physics, we performed DFT+$U$ calculations with $U$=3.0 and 4.0 eV, as well as calculations using the hybrid HSE06 functional. For each case, we evaluated the $m_e^*$ and $n_e$ for both $C_3$-symmetric FM and AB-AFM spin configurations. The results are summarized in Table S2. Although $m_e^*$ and $n_e$ vary moderately across different functionals and spin orderings, the calculated $r_s$ remains consistently large in all cases, sufficient to support the formation of a Wigner crystal.

**Possible explanations of the charge-density modulations**

Possibilities beyond Wigner crystal for interpreting the observed charge modulations mainly include QPI and Friedel oscillation i) QPI. Theoretically, QPI, arising from the interference of elastically scattered quasiparticles on constant-energy contours, would exhibit an energy dispersion that reflects the underlying electronic structure[67,68]. The energy independence of Wigner crystal II across a large range (up to 2.8 eV) (Fig. 2i) excludes the QPI-based explanation. ii) Friedel oscillation. In principle, VCl$_3$ as an insulator cannot exhibit the Friedel oscillation. Although charge carriers that enable the Friedel oscillations can be generated if the tip-induced band bending is sufficiently strong, the resulting period of oscillation would depend on the density of the induced carriers, thus on the bias voltage. This is apparently inconsistent with the bias-independent behavior of Wigner crystal II observed in our experiments. Charge Friedel oscillation in a Mott insulator can also occur in the presence of a spinon Fermi surface[69] in the spin liquid phase[70], which can yield a spatial static ordering. Yet, VCl$_3$ is magnetic insulator at $T$<21.8 K[33] instead of a spin liquid, precluding this spinon-related scenario.

**Nematicity in the Wigner crystal**

We discuss the interpretations of the nematicity manifested in Wigner crystal I. i) Landau level mixing. In 2D electron systems, e.g., GaAs/(AlGa)As quantum well[3,4], the partially filled Landau level can host intricate competition among fractional quantum Hall liquid, Wigner crystal and isospin ferromagnets[39,71-73]. The highest Landau level, proposed to be alternately full or empty in space[40,74,75], yields intriguing intermediate stripe and bubble Wigner crystals[25,40,76]. Yet, in this picture, the expected $C_2$-symmetric stripe, appearing whenever a Landau level is filled, contrasts with our nematicity that emerges without the need of magnetic field, and meanwhile evolves monotonically as increasing the magnetic field (Fig. 3s). ii) Breaking of $d_{xz}/d_{yz}$ degeneracy. The splitting of $d_{xz}/d_{yz}$ orbitals has been reported to explain electronic nematicity in iron-based superconductors[77], which, however, remains largely insensitive to an out-of-plane magnetic field. Moreover, the parent phase with quenched nematicity (i.e. degenerate $d_{xz}/d_{yz}$ orbitals) retains a $C_4$ symmetry, rather than the $C_6$ symmetry relevant in the present case (e.g. Fig. 3g).

The exclusion of available possibilities, e.g., Landau level mixing and lifting of $d_{xz}/d_{yz}$ degeneracy, suggests that the spontaneous in-plane stripped spin correlation of Wigner solid is likely appropriate for interpreting the nematicity here. Recently, modified variational Monte Carlo calculations, which offer improved accuracy, reveal that the nematic AFM spin stripes are energetically favored in Wigner crystals with intermediate charge densities ($r_s$~10–35)[45]. The predicted spin nematicity is electronically driven, not materials specific, in agreement with our Wigner crystal I with flexible nematic orientations that are unrelated to the structural trimer lattice of VCl$_3$. In addition, under the influence of strong spin–orbit coupling, the electronic structures can be modified depending on the spin orientation[78,79]. The modification can be realized, for example, through lifting the $3d$-band degeneracies[80], or hybridization of different minority orbital bands, both of which are sensitive to magnetization direction[78]. We thus expect that the preferred direction of the more-intense Wigner crystal would modulate substantially with the orientation of striped in-plane spin-correlation phase. Such coupling makes the spin-modulated Wigner crystal observable even with a non–spin-resolved STM as in our experiments.

**Distinct itinerancy for the transferred electrons**

The difference in itinerancy for the transferred electrons can be partly understood through a polaron-enhanced mechanism. In this picture, electron–phonon coupling increases the effective mass of the lower branch of 4×4 supercell-induced subband, and opens a global gap relative to the higher subbands. Both effects stabilize the lower subband that contributes to Wigner crystal I with high $T_c$. In contrast, the higher subband is more dispersive, because: i) it contains only a fractional number (~0.3) of electrons per supercell, and thus cannot fully occupy the band, ii) it experiences repulsive interactions from the localized electrons in the lower subband. Consequently, the Wigner crystal II that consists of the higher subband electrons manifests as a lower-$T_c$, more liquid-like state. Such itinerancy difference is also hinted at their disparate tunneling characters. For example, the nematic Wigner crystal I is detectable at energies within the VCl$_3$ gap, while not for Wigner crystal II under the same conditions (Fig. S7). This tunneling-absent behavior can be attributed to the stronger itinerancy of Wigner crystal II, which screens the tunneling electrons from the graphite substrate, thereby suppressing any tunneling current, especially the one within VCl$_3$ gap where the signal is already sufficiently weak.

The spatial distribution of the transferred electrons in VCl$_3$ would be influenced by the electron puddle-related inhomogeneity[81], resulting in a spatially varying density for the 'polaron-trapped' electrons of Wigner crystal I. Such variation naturally explains the observed diversity in the period and orientation of Wigner crystal I. Besides the puddle-type inhomogeneity, the 'polaron-trapped' electron density can also be susceptible to many other factors, such as growth conditions (annealing time, temperature, etc.), layer thickness of the decoupled graphene on graphite, interface coupling strength, and the twist angle between VCl$_3$ and graphite. In addition, it should be noted that the DFT-based calculations generally capture short-range couplings accurately, but provide only a limited description of the long-range interactions. The lower subband may therefore be overly influenced by short-range electron–phonon coupling, leading to excessively localized small polarons, as seen in our DFT+$U$ results (Fig. 5b). In reality, when the long-range couplings are fully considered, the spatial extent of the polaronic states associated with the lower subband is expected to be much broader as experimentally imaged.

## Data availability

The data that support the findings of this study are available from the corresponding authors upon reasonable request.

## Acknowledgements

The authors acknowledge Fengcheng Wu for helpful discussions. This work is funded by the National Key Research and Development Program of China (Grant No. 2022YFA1402400), the National Science Foundation of China (Grants No. 92265201, 92477137, 12174131, 12404210, and 12574194), and Science Challenge Project (Grant No. TZ2025013).

## Author contributions

Q.-K.X. and Y.-S.F. supervised the project. J.Z. and W.L. grew the sample and conducted the STM experiments under assistance from C.Z. and T.G. and guidance from Y.-S.F., C.L., and W. Z. Z.J. carried out first-principles calculations. J.Z. and W.L. performed the data analysis and figure development supervised by Y.-S.F. and C.L. C.L. and Y.-S.F. wrote the paper with contributions from all authors.

## Competing interests
The authors declare no competing interests.

## Additional information
Correspondence and requests for materials should be addressed to Ying-Shuang Fu, or Qi-Kun Xue.